\documentclass[12pt]{article}

\usepackage{amsmath,amssymb,amsthm,units,stmaryrd,upgreek}
\usepackage[usenames,dvipsnames]{xcolor}
\usepackage{yfonts}
\usepackage{units,stmaryrd}
\usepackage{color,xspace}
\usepackage{graphicx}
\usepackage[all]{xy}

\newtheorem{definition}{Definition}[section]
\newtheorem{lemm}{Lemma}[section]

\newtheorem{theorem}{Theorem}[section]
\newtheorem{example}{Example}[section]
\usepackage{bm}

\newcommand{\dfadd}[1]{{#1}}
\newcommand\para[1]{\medskip \noindent {\emph{#1.}}}

\newcommand\Comma{\mathord,}

\newcommand\End{{\tt end}}
\newcommand\pass{{\tt pass}}

\newcommand\sadi{\ensuremath{\rm SADI}\xspace}

\newcommand\deck{\ensuremath{\mathsf{Deck}\xspace}} 
\newcommand\deal[1]{\mathop{\rm Deal}(#1)}
\newcommand\cardsum[1]{|#1|}
\newcommand\players{\ensuremath{\mathsf{Agt}\xspace}} 
\newcommand\actions{\mathbb A}
\newcommand\distr{{\rm Dist}}
\newcommand\distfun[1]{\| #1\|}
\newcommand\eavesig[2]{{\mathcal I}_{#2}(#1)}

\newcommand{\vcut}[1]{}

\newcommand{\dnote}[1]{}
\newcommand{\vfnote}[1]{}
\usepackage[colorinlistoftodos,textwidth=100pt]{todonotes}
\newcommand{\vnote}[1]{\todo[color=green!50!white]{\ \\\color{black}\bf{}{VG: #1}}}
\newcommand\vgadd[1]{{#1}}

\newcommand{\ali}{\ensuremath{\mathcal{A}}\xspace} 
\newcommand{\bob}{\ensuremath{\mathcal{B}}\xspace} 
\newcommand{\cat}{\ensuremath{\mathcal{C}}\xspace} 
\newcommand{\eaves}{\ensuremath{\mathcal{E}}\xspace} 
\newcommand{\spread}{spread\xspace}
\newcommand{\dist}{\ensuremath{\bar s}\xspace}
\renewcommand{\actions}{\ensuremath{\mathsf{Act}\xspace}}

\newcommand{\run}{\ensuremath{\rho}\xspace} 
\newcommand{\prot}{\ensuremath{\pi}\xspace} 

\newcommand{\prsadi}{\ensuremath{\Sigma}\xspace} 

\begin{document}

\author{David Fern\'{a}ndez-Duque\\
Department of Mathematics\\
Instituto Tecnol\'ogico Aut\'onomo de M\'exico\\
R\'io Hondo 1, 01080 Mexico City, Mexico\\
{\tt david.fernandez@itam.mx}\\\\
Valentin Goranko\\    
Department of Philosophy\\
Stockholm University\\
SE - 10691 Stockholm, Sweden\\
{\tt valentin.goranko@philosophy.su.se}
}

\title{Secure aggregation of distributed information:\\
\large How a team of agents can safely share secrets in front of a spy}

\maketitle


\begin{abstract}
We consider the generic problem of Secure Aggregation of Distributed Information (SADI), where several agents acting as a team have information distributed amongst them, modelled by means of a publicly known deck of cards distributed amongst the agents, so that each of them knows only her cards. The agents have to exchange and aggregate the information 
about how the cards are distributed amongst them by means of public announcements over insecure communication channels, intercepted by an adversary ``eavesdropper'', in such a way that the adversary does not learn who holds any of the cards.  
%
We present a combinatorial construction of protocols that provides a direct solution of a class of SADI problems and develop a technique of iterated reduction of SADI problems to smaller ones which are eventually solvable directly. We show that our methods provide a solution to a large class of SADI problems, {including all SADI problems with sufficiently large size and sufficiently balanced card distributions.} 
\end{abstract}

\section{Introduction}

We consider a generic scenario where a set of agents $\players$ have information distributed amongst them, i.e., included in their collective knowledge, while each agent has only partial knowledge of it. The agents act as a team that has to exchange and aggregate that information, either as common knowledge within their group or in the individual knowledge of at last one of them. The exchange is performed over insecure communication channels and is presumed intercepted by an adversary. The task of the team is to achieve the aggregation of the distributed information, following a prearranged protocol, in such a way that the adversary does not learn important information. 

More specifically, we model the problem by assuming that the information that each agent has is encoded by a set of ``cards'' that she\footnote{For convenience of exposition, we will assume that the agents are female while the eavesdropper is male.} holds in her  hands, where the cards are drawn from a publicly known deck\footnote{The drawing and distribution of these cards is considered secret and  secure and we will not discuss the side issue of how exactly that is done. In reality, we assume that each of the agents has obtained her initial information in some private way.} and every card is in the hands of exactly one agent of the team. \dfadd{The deck of cards should be seen merely as a metaphor for the information held by each agent, an idea that has proven useful for modelling secure computations in several settings unrelated to our own \cite{MizukiS14,NiemiR97,Stiglic01}.} The goal of the team is to exchange and disseminate across the whole team the information about how the cards are distributed among the agents. It is assumed that the agents can only communicate by making public announcements over insecure channels and that there is an ``eavesdropper'' Eaves ($\eaves$) whose goal is to learn as much as possible about the distribution of the cards by intercepting and analysing the announcements exchanged by the agents in $\players$. In particular, Eaves wants to learn who owns at least one of the cards. We further assume that in their exchange of announcements the agents follow a publicly known (hence, known by the  eavesdropper, too) protocol. 

The scenario described above is a variation of the well-known ``Russian cards problem'', which is more than one-and-a-half centuries old \cite{kirkman:1847} but has recently had renewed attention \cite{hvd.studlog:2003}, leading to many new solutions (e.g. \cite{albertetal:2005,geometric,swanson:2012,swanson:2014}). Here we will generalize the problem substantially by allowing an arbitrary number of agents, but on the other hand we restrict it essentially by assuming that the eavesdropper has no cards in his hands\footnote{\vgadd{The effect of allocating cards to the eavesdropper is deeper than just the fact that not all cards are in the hands of the team. It also creates the danger that the announcements of the  agents in the team about cards they do not hold may reveal unwanted information to the eavesdropper. So, the solution protocols developed here would generally not work in the case where the eavesdropper holds cards, and we leave that case for future work.}}. \vgadd{According to our knowledge}, such a multi-agent setup had only previously been considered in \cite{Duan}, although our approach is very different. Interest in this problem arises from the fact that it is based on {\em information-theoretic cryptography} \cite{maurer:1999}, where security is not contingent on the computational complexity of breaking the code but rather on communications that do not contain sufficient information for an eavesdropper to learn the original message. 

\subsection*{Main results and contributions:}


{In this paper, we introduce the generic {\em Secure Aggregation of Distributed Information (SADI)}  problem and model it in the style of the Russian cards problem. We introduce a formal framework for specifying SADI problems involving any number of communicating agents and leading to several notions of security and informativity. We then focus on a version of SADI problems with natural safety and informativity conditions, for which we present a combinatorial construction of protocols that provides a direct solution of a class of SADI problems and then develop a general technique for solving the problem by reducing it recursively to smaller instances. Finally, we show how this method can be used to solve a wide class of SADI problems, including all SADI problems with sufficiently large size and sufficiently balanced card distribution.

Our results and methods \dfadd{may eventually} be used for developing practical protocols for secure communication, which we briefly suggest in the concluding section.}

\subsection*{Organization of the paper:} 

We motivate the current work in Section \ref{sec:illustr} by presenting a detailed example which showcases some of the notions that will arise throughout the text. Section \ref{sec.SADI} then provides the general setup of the Secure Aggregation of Distributed Information (\sadi) problem. In Section \ref{sect:3agents} we focus on solving the \sadi problem in the 3-agent case, and in Section \ref{SecCaseMult} we set the stage for working with more agents. Section \ref{sec:reduction:general} describes a general technique by reduction to smaller cases, which is then employed in Section \ref{SecSolvability} to prove that a large class of instances of the \sadi problem are solvable.
{In a brief concluding section we suggest further extensions of our techniques and some applications. Then, we include in an appendix some more technical proofs consisting of algebraic manipulations.} 

\section{An illustrative example}
\label{sec:illustr}

Before we present the generic setup and embark on a general analysis of the multi-agent setting, we begin with a non-trivial illustrative example of the type of problems we consider in the paper. 
It involves a team of three agents\footnote{Note that the case of two agents that hold all the cards is trivial as they know the distribution from the beginning.}, Alice (\ali), Bob (\bob) and Cath (\cat) who  hold respectively $2,3$ and $4$ cards, identified with the numbers $1,\hdots,9$. 

We are interested in designing a protocol that would eventually inform each of the agents about the deal, while the eavesdropper Eaves (\eaves) may not learn the ownership of any of the cards. 
\vcut{Formally, we consider the \sadi problem $\big((2,3,4),\text{\sc i},\text{\sc s}\big)$.}

We will describe informally a protocol solving this problem, by describing it on a (randomly chosen) particular deal in which we assume, without loss of generality, that Alice gets $\{1,2\}$, Bob gets $\{3,4,5\}$, and Cath gets the remaining cards $\{6,7,8,9\}$. We will use the notation $H_\ali|H_\bob|H_\cat$ to represent the deal and may omit set-brackets, so that the deal may also be written as $1,2\mid 3,4,5\mid  6,7,8,9$.

\para{Step 1}
Alice chooses at random a card not in her hand, say 9. Then she makes an 
announcement, saying (essentially):
\begin{quote}
\textit{``My cards are among $\{1,2,9\}$''.}
\end{quote}
After such announcement, the agent who holds the extra card (9) -- in this case Cath -- knows the card distribution.  

\para{Step 2} That agent (Cath) makes the next announcement, which has to inform the others of the distribution, as follows. There are three possible ways that the cards $1,2,9$ may be distributed among Alice and Cath: $1,2\mid 9$, \ $2, 9\mid 1$ or $1,9\mid 2$. 

Note that Alice's
hand in this context is determined by Cath's card within $\{1,2,9\}$, so we may represent the three possibilities by Cath's card, and these form a set $\Gamma=\big\{1,2,9\}$. 
Once Alice's cards are known, the rest of the deal is determined by Bob's hand. There are many hands that Bob may hold which are consistent with Alice's announcement: $\{3,4,5\}$ (his actual hand), but also, for example, $\{5,6,7\}$, etc. Let $\Delta$ be the set of all such hands.

Cath will then choose a map $f\colon \Gamma\to \Delta$ such that:
\begin{enumerate}
\item All cards are mentioned in the domain or range of $f$ (else Eaves will learn some of Alice's cards). 
\item No card belongs to all values of the mapping 
 (else Eaves would learn that the card is in Bob's hand). 
\item The mapping is injective  
 (but not necessarily onto).
\item Cath's actual card is mapped to Bob's actual hand 
(so that both Alice and Bob can learn the distribution after that announcement).  
\item All other values of the mapping are chosen at random  
(so that Eaves cannot learn more than intended from the protocol). 
\end{enumerate}

One such mapping is
\begin{align*}
f(\cat: 9) &= \bob:\{3,4,5\}, \\ 
f(\cat: 2) &= \bob:\{5,6,7\}, \\ 
f(\cat:1) &= \bob:\{6,7,8\}.     
\end{align*}

This mapping in turn gives rise to a set of possible deals; for example, if Cath has $9$ Alice has $\{1,2\}$, and according to $f$, Bob should have $\{3,4,5\}$, so that Cath should hold the remaining cards.  

Now, Cath announces that
\begin{quote}
\textit{``The actual deal belongs to the set}
\begin{equation}\label{EqCathDiff}
\big \{1\Comma 2|3\Comma 4\Comma 5| 6\Comma 7\Comma 8\Comma 9;  \ 1\Comma 9| 5\Comma 6\Comma 7| 2\Comma 3\Comma 4\Comma 8;\ 2\Comma 9| 6\Comma 7\Comma 8 | 1\Comma 3\Comma 4\Comma 5\big\}.\textit{''}\end{equation}
\end{quote}

This announcement completes the protocol.

\medskip
We claim two important properties of the protocol presented above, which we leave the reader to check: 

\begin{enumerate}
\item It is  \emph{informative} for all agents, in the sense that they all eventually learn the card distribution.  
\item It is  \emph{card-safe} in the sense that the eavesdropper does not learn the ownership of any of the 9 cards.
\end{enumerate}

This example gives the basic intuition behind the protocols we will work with. Before considering a more general setting, we formally define the concepts of informative and safe protocols in the next section.

\section{Secure Aggregation of Distributed Information Problems} 
 \label{sec.SADI}

Here we will give precise definitions needed to set up the information aggregation problem. If $X$ is a set and $n$ a natural number, we use $\binom Xn$ to denote the subsets of $X$ of cardinality $n$. The cardinality of $X$ is denoted $\# X$.

\subsection{Basic terminology and notation}
\label{SubsecAnnRunProt}

\begin{definition}
Let $\players$ be a finite set of agents (or `players'). By a {\em distribution type} we mean a vector $\dist =(s_P)_{P\in \players}$ of natural numbers. We write $\cardsum {\dist}$ for $\sum_{P\in\players}s_P$.

The deck, $\deck$, is a finite set of cards with cardinality $\cardsum {\dist}$. When not mentioned explicitly we assume that $\deck= \{1,\hdots,\cardsum{\dist}\}$. A  {\em deal} of type $\dist$ over $\deck$ is a partition $H=(H_P)_{P\in \players}$ of $\deck$ such that $|H_P|=s_P$ for each agent $P$. We say $H_P$ is the {\em hand} of $P$. We denote the set of all deals of type  $\dist$  over $\deck$ by $\deal{\dist,\deck}$, or merely $\deal{\dist}$ if $\deck=\{1,\hdots,\cardsum{\dist}\}$.

If $H$ is a deal, we denote by $\distfun H$ its distribution type, i.e. $\|H\|_P=\#H_P$ for each agent $P$.
\end{definition}

As noted earlier, we consider that there is an initial secure dealing phase in which a card deal is selected randomly. The process by which the cards are distributed is treated as a black box. Afterwards, the agents have knowledge of their own cards and of the distribution type $\dist$ of the deal, but know nothing more about others' cards. Thus, they are not able to distinguish between different deals where they hold the same hand. We model this by equivalence relations between deals; since from the perspective of agent $P$, a deal $H$ is indistinguishable from deal $H'$ whenever $H_P=H'_P$, we define $H\sim_P H'$ if and only if $H_P=H'_P$. If the agents are numbered $P_1,\hdots,P_m$, we may write $\sim_i$ instead of $\sim_{P_i}$.

In \cite{albertetal:2005,swanson:2012} and elsewhere, an action has been modelled as an announcement of a set of hands that one of the agents may hold. Thus, the agent Alice ($\ali$) would announce a subset $\mathcal S$ of $\deck\choose a$, indicating that $H_\ali\in\mathcal S$. In our setting, however, announcing information about one's own hand may not be enough, as an agent may wish to share knowledge they have about the rest of the deal. Thus, a general form of an announcement will be a set of deals $\mathcal S\subseteq\deal{\dist,\deck}$.\footnote{{Agents may also be allowed to make announcements which are not precisely of this form. As we will see later, such announcements can usually be simulated by announcing, instead, the set of deals for which the announcement would be true.}}
 Moreover, given that there are now more agents, the amount of actions needed to distribute the information may vary. Because of this, we will add an additional action, $\End$, whose sole purpose is to stop communications once the goals have been achieved. For our information protocols we will assume throughout that agents take turns, so that if the agents are listed by $P_1,\hdots,P_m$, then $P_1$ realizes an action first, followed by $P_2$, etc. Note that this contrasts with the example in Section \ref{sec:illustr} where Alice goes first, followed by Cath; we may accommodate for this by allowing Bob to ``pass''.
This can be modeled by making vacuous announcements, 
  to be made precise later (see Subsection \ref{SubsecTypesAnn}). 

In the presentation of protocols we will closely follow that in \cite{colouring}.

\begin{definition}[Runs]\label{def:run}
Let $\actions=\mathcal P(\deal{\dist,\deck})\cup \{\End\}$. 
The elements of $\actions$ will be called {\em actions}. 
A {\em (finite) run} is a (possibly empty) sequence $\run=\alpha_1,\hdots,\alpha_n$ of actions from $\actions$. The empty run is denoted by $()$. If $\run=\alpha_1,\hdots,\alpha_n$ and $\alpha$ is an action we write $\rho\ast \alpha$ for $\alpha_1,\hdots,\alpha_n,\alpha$. An {\em infinite run} is an infinite sequence $\alpha_0,\alpha_1,\alpha_2,\hdots$ of actions. Runs will be assumed finite unless it is explicitly stated otherwise. We denote the length of a run $\rho$ by $|\rho|$.

A run is  {\em terminal} if its last action is $\End$. A run is {\em proper} if it contains no occurrences of $\End$ except possibly for the last action. We denote the set of proper runs by ${\rm Run}$.

For a run $\run=\alpha_1,\hdots,\alpha_n$, let $\bigcap \run$ denote the set
\[\bigcap\{\alpha_i : {1\leq i\leq n \text{ and } \alpha_i\not=\End}\}.\]
\end{definition}

We now define the notion of {\em protocol} we will use. Below and throughout the text, we use $(x)_{d}$ to mean the unique $r\in [1,d]$ such that $x\equiv r\pmod d$. This notation will be a useful shorthand to indicate the player whose turn it is after $x$ steps.

\begin{definition}[Protocol]\label{defprot}
Let ${\rm Deal}=\deal{\dist}$.

A {\em protocol} (for $\dist$) is a function $\pi$ assigning to every deal $H\in{\rm Deal}$ and every non-terminal proper run $\rho\in{\rm Run}$ a non-empty set of actions $\pi(H,\rho)\subseteq \actions$ such that if 
$\alpha \neq \End$ and $\alpha\in \pi(H,\rho)$ then $H\in \alpha$ and if $i=(|\rho|+1)_m$ (so that it is the turn of the agent $P_i$) and $H\sim_{i}H'$ then $\pi(H,\rho)=\pi(H', \rho)$.

An {\em execution of a protocol $\pi$} is a pair $(H,\run)$ of a deal $H\in{\rm Deal}$ and a run $\run=\alpha_1,\hdots,\alpha_n$, such that $\alpha_{i+1} \in \pi(H,\rho[1..i])$ for every $i<n$, where $\rho[1..i] = \alpha_1,\hdots,\alpha_i$.


An execution of a protocol $(H,\run)$ is {\em terminating} if the run $\run$ is terminating, i.e. if its last element is $\End$. A protocol is {\em terminating} if it has no infinite executions.
\end{definition}

Thus, a protocol is a tree-like set of runs representing a non-deterministic strategy for the communicating agents. Once a deal has been fixed, a protocol assigns to each run a set of actions out of which the agent whose turn it is must choose one at random. These actions are determined exclusively by the information the agent who is to move has access to, which is assumed to be {\em only:} ({\it i}) her hand, ({\it ii}) the distribution type  $\dist$ of the deck $\deck$, ({\it iii}) the announcements that have been made previously and ({\it iv}) the protocol being executed.   
Note that protocols are generally non-deterministic and hence may have many executions.

\subsection{Some useful types of announcements}
\label{SubsecTypesAnn}

Since we will often be using announcements of a very particular type, it will be convenient to provide a more compact notation for them.

\para{1}  An agent $P$ may merely announce a set of hands $\mathcal S\subseteq {{\deck}\choose {s_P}}$ such that $H_P\in\mathcal S$. This announcement can be modeled as a set of deals, namely
\[\{H'\in \deal{\dist}: H'_P\in\mathcal S\}.\]

\para{2} Let $S$ be a set of cards and $P$ an agent, and suppose that $P$ holds $n$ cards in $S$, that is, $\#(H_P\cap S)=n$. She may then wish to announce ``I hold $n$ cards in $S$.'' This may also be represented as a set of deals, namely
\[\{H'\in \deal{\dist}:\#(H'_P\cap S)=n\}.\]
An important special case is the one where $H_P\subseteq S$, in which case the agent may state ``All my cards are among $S$''.

\para{3} The agent $P$ may also announce a set of {\em restricted} deals. To be precise, if $B\subseteq\deck$ and $H$ is any deal, let $H'=H\upharpoonright B$ denote a deal over the deck $B$ such that $H'_P=H_P\cap B$ for each agent $P$, and let $\bar t=\|H'\|$. Then, the agent may announce ``The deal restricted to $B$ belongs to $\mathcal S\subseteq\deal{\bar t,B}$''. This corresponds to announcing the set of deals
\[\{H\in\deal{\dist}:H\upharpoonright B\in \mathcal S\}.\]
Note that for such an announcement we assume that $P$ already knows the distribution $\bar t$, usually as a result of others having announced how many cards they hold in $B$.

\para{4} Agents may choose to ``pass''. This may be modeled by them simply announcing all of $\deal{{\dist}}$ (as such an announcement contains no factual information). We will denote this announcement by $\pass$.

\medskip
Note that when an agent announces ``I hold $n$ cards in $S$,'' she does not explicitly mention $n$ or $S$ since our announcements are {\em only} sets of deals. As such announcements play a prominent role in our protocols, it will be useful to show that other agents can essentially infer the values of $n$ and $S$, which is the meaning of the next lemma.

\begin{lemm}\label{LemmCountAnn}
Let $\dist$ be any distribution type, $\alpha$ be the announcement ``I hold $n$ cards in $S$'' by agent $P$, where $\#S>n$ and $s_P\geq n$ and $s_P-n<\#(\deck\setminus S)$, and let $\beta$ be the announcement ``I hold $m$ cards in $T$'' with $m<\# T$. Then, $\alpha=\beta$ if and only if either $m=n$ and $T=S$ or $m=s_P-n$ and $T=\deck\setminus S$.
\end{lemm}

{
\proof
Clearly, $\alpha=\beta$ if $m=n$ and $T=S$, or $m=s_P-n$ and $T=\deck\setminus S$.

Conversely, assume that $\alpha=\beta$. First we note that if $S=T$ then $n=m$ since for any $H\in \alpha$, $\# (H_P\cap S)=n$, so $n$ is uniquely determined by $\# (H_P\cap S)$. Similarly, if $T = Deck\setminus S$ then $m = 
s_P -n$, since for any $H\in \alpha$, $\# (H_P\cap T)=\# H_P-\# (H_P\cap S)=s_P-n$.

Therefore, toward a contradiction we may assume that $T\not=S$ and also $T\not=(\deck\setminus S)$, and consider three cases. 

\para 1 If $T\subsetneq S$, let $x\in S\setminus T$ and $A\subseteq S\setminus \{x\}$ 
be arbitrary with $n-1$ elements. Further, let $B\subsetneq \deck\setminus S$ be arbitrary with $s_P-n$ elements and $y\in \deck\setminus(S\cup B)$ be arbitrary (note that our inequalities guarantee that all these conditions can be met). Consider a deal $H$ where $H_P=A\cup B\cup \{x\}$ and all other hands chosen randomly.  
Consider also a deal $H'$ with $H'_P=A\cup B\cup \{y\}$ and all other hands chosen randomly as well. Clearly, $\#(H_P\cap S)=n$ so $H\in \alpha$, but $\#(H'_P\cap S)=n-1$, so $H'\not\in\alpha$. Since we are assuming that $\alpha=\beta$, we also have $H\in \beta$, but since $x,y\not\in T$, $\#(H'_P\cap T)=\#(H_P\cap T)=m$ which would imply that $H'\in \beta$ and thus $\alpha\not=\beta$, a contradiction. 

\para 2 For the case where $T$ is disjoint from $S$ we may replace $S$ by $\deck\setminus S$ and proceed as above, noting that $T\subsetneq \deck\setminus S$.

\para 3 Finally, we are left with the case where neither $T\subsetneq S$ nor $T$ is disjoint from $S$. Thus there are $x,y$ with $x\in S\cap T$ and $y\in T\setminus S$. Let $A\subseteq S\setminus \{x\}$ be an arbitrary set with $n-1$ elements, $B\subseteq \deck\setminus(S\cup \{y\})$ have $s_P-n$ elements, and consider two deals $H,H'$, where $H_P=A\cup B\cup \{x\}$ and $H'_P=A\cup B\cup \{y\}$. Then, $P$ holds $n$ cards from $S$ in $H_P$, so that $H\in \alpha$; but $\#(H_P\cap T)=\#(H'_P\cap T)$, so that $H\in \beta$ implies that $H'\in\beta$. However, $H'_P\cap S$ has $n-1$ elements and thus $H'\not\in \alpha$, so that $\alpha\not=\beta$, a contradiction.
\endproof
}

\subsection{Informative and safe protocols. \sadi problems}
\label{SubsecTypesAnn}

Now we will define two important properties of protocols in terms of which we will formulate the type of problems studied in our setting. The first property is \emph{informativity}: that agents in the team learn some or all of each other's cards (or,  the entire deal) at the end of its execution:

\begin{definition}[Informativity]
An execution $(H,\run)$ of a protocol $\prot$ is  {\em informative for an agent $P$} if there is no execution $(H' ,\run)$ of $\prot$ with $H'\not=H$ but $H_P=H'_P$ (i.e., at the end of the run the agent knows the precise card distribution.) 

A terminating protocol $\prot$ is
\begin{itemize}
\item[{\sc wi:}] {\em weakly informative} if every terminating execution of $\prot$ is informative for \emph{some} agent in $\players$.
\item[{\sc i:}] {\em informative} if every terminating execution of $\prot$ is informative for \emph{every} agent in $\players$.

\end{itemize} 
\end{definition}

Clearly, {\sc i} implies {\sc wi}, and in general they are not equivalent. Note that the proof that a given protocol is informative can be assumed to be common knowledge among the agents, and therefore the distribution of the cards at the end of every execution becomes their common knowledge, too.

The second important property is \emph{safety}: for any card $c$, the eavesdropper Eaves should not know who holds it. To formulate Safety, let us first define the eavesdropper's {\em ignorance set.}

\begin{definition}
Given a protocol $\prot$ and a run $\run$, define the {\em (eavesdropper's) ignorance set} $\eavesig\run\prot$ as the set of all deals $H$ such that $(H,\run)$ is an execution of $\prot$.
\end{definition}

Thus, Eaves cannot rule out any deal in $\eavesig\run\prot$ even if he has full knowledge of the protocol and all announcements in $\rho$ have been made. We use this to formalize our notions of safety, which require that Eaves not be able to determine the ownership of some or all cards or of the entire deal.

\begin{definition}[Safety of cards]\label{def:cardsafe}
An execution $(H,\run)$ of a protocol $\prot$ is {\em safe for the card $c$} if for every agent $P$, if $c \in H_P$ there is $H'\in\eavesig \run\prot$ such that $c\not\in H'_P$. 
It is {\em strongly safe for the card $c$} if for every agent $P$, there is $H'\in\eavesig \run\prot$ such that $c\in H'_P$ and there is $H''\in\eavesig \run\prot$ such that $c\not\in H''_P$.
\end{definition}

Note that it is not enough for isolated runs to be safe, however; since we are interested in unconditionally secure protocols, we require for {\em every} execution of a protocol to be safe.

\begin{definition}[Safety of protocols]\label{def:protsafe}
A protocol $\prot$ is: 
\begin{itemize}
\item[{\sc ds:}] {\em deal-safe} if every execution of $\prot$ is safe for \emph{some} card $c$. Equivalently, deal-safe means that  the eavesdropper does not learn the deal at the end of any execution of $\prot$. 

\item[{\sc s$_P$:}] {\em $P$-safe}, for an agent $P$, if every execution of $\prot$ is safe for \emph{all} cards in $H_P$. 

\item[{\sc s:}] {\em (card-)safe} if every execution of $\prot$ is safe for \emph{every} card $c$. 

\item[{\sc ss:}] {\em strongly (card-)safe} if every execution of $\prot$ is strongly safe for \emph{every} card $c$.
\end{itemize}
\end{definition}

Once again we list these conditions from weakest to strongest, so that {\sc ss} implies {\sc s}, which implies {\sc s$_{P}$} for any player $P$, which in turn implies {\sc ds}. With card-safe protocols the opponent never learns any \emph{positive} information about the ownership of any card, but he may learn negative information about non-ownership of cards. With strongly card-safe protocols the opponent learns neither positive nor negative information about the ownership of any card.

Now, we can define the general type of problems we are interested in.
 
\begin{definition}[\sadi problems]
\label{def:sadip}
A \emph{Secure Aggregation of Distributed Information Problem (\sadi)} 
 is a triple $(\dist,\iota,\sigma)$  consisting of a distribution type $\dist$, an  informativity condition $\iota\in\{\text{\sc wi,i}\}$ and a safety condition $\sigma\in \{\text{\sc ds,s$_P$,s,ss}\}$. 
\end{definition}

\begin{definition}[Solvable \sadi problems]\label{DefSolSadi}
A  \sadi problem $(\dist,\iota,\sigma)$ is {\em solvable} if there exists a terminating protocol $\pi$ for $\dist$ that satisfies the safety condition $\iota$ and the informativity condition $\sigma$. Every such protocol is called a  {\em solution} of the \sadi problem.  
\end{definition}

In this paper we will focus on the case of safe and informative protocols, i.e. $\iota=\text{\sc i}$ and $\sigma=\text{\sc s}$. Hereafter, by a \sadi problem we will mean one of this type.

\section{Informative and safe protocols for the three-agent case}
\label{sect:3agents}

In Section \ref{sec:illustr} we considered the \sadi problem $\big((2,3,4),\text{\sc i},\text{\sc s}\big)$. Now, we are going to consider the general three-agent case and to obtain a generic solution under some simple sufficient conditions. Before describing that solution, we need some technical preparation.

\subsection{Spreads}
\label{SubsecSpread}

Let us now introduce {\em spreads,} a technical notion that generalizes the type of announcement completing the protocol in the case of $\big((2,3,4),\text{\sc i},\text{\sc s}\big)$.

\begin{definition}[(basic) \spread]  
Let $Y,Z$ be sets and $\binom Z n$ be the set of subsets of $Z$ of cardinality $n$. 
A mapping $f\colon Y \rightarrow \binom {Z}{n}$ is a \emph{\spread} iff:

\begin{enumerate}
\item (Injection) \ $f$ is injective;
\item (Coverage) \ $\bigcup_{y\in Y}f(y)= Z$; 
\item (Avoidance) \ $\bigcap_{y\in Y}f(y)=\varnothing$.
\end{enumerate}
\end{definition}

\vcut{
\begin{lemm}
\label{lem:spread}
Let $|Y| = \textbf{y}$, $|Z| = \textbf{z}$. Then a \spread  $f: Y \rightarrow \binom {Z}{n}$ exists if and only if the following conditions hold: 
\begin{enumerate}
\item (Injection) \ $\binom {z}{n} \geq y$.
\item (Coverage) \ $ny \geq z$.
\item (Avoidance) \ $(y-1)z \geq ny$. 
\end{enumerate}
\end{lemm}
 }

\begin{lemm}
\label{lem:spread}
Let $|Y| = k$, $|Z| = m$. Then a \spread  $f: Y \rightarrow \binom {Z}{n}$ exists if and only if the following conditions hold: 
\begin{enumerate}
\item (Injection) \ $\binom {m}{n} \geq k$.
\item (Coverage) \ $nk \geq m$.
\item (Avoidance) \ $(k-1)m \geq nk$. 
\end{enumerate}
\end{lemm}

\proof {The necessity of each of the first two conditions is straightforward. For Avoidance, let $Y = \{y_{1},\ldots,y_{k} \}$. Then each of the elements in $Z$ must appear at most $k-1$ times in all $n$-element sets $f(y_{1}),\ldots,f(y_{k})$, so there must be a total of at most $m(k-1)$ occurrences of elements of $Z$ in these sets. On the other hand, the number of these occurrences is $nk$, whence the inequality. 

Conversely, if all three conditions are satisfied, then a spread can be constructed as follows. Without loss of generality we may assume that $Z=\{1,2,\hdots,m\}$, and likewise $Y=\{1,2,\hdots,k\}$. Suppose first that $2n\leq m$, and let $q=\lceil \nicefrac m n \rceil$. Observe that from $nk\geq m$ we obtain $q\leq k$. Then, for $1\leq i\leq q$, we define
\[f(i)=\{((i-1)n+j)_m\colon 1\leq j\leq n\}.\]
It should be clear that if $i\not=i'$ then $f(i)\not=f(i')$. For $i>q$, choose $f(i)\in {Z\choose n}$ at random in such a way that $f$ is injective; this may be obtained in view of the Injection condition.

It remains to check the other two conditions, but Coverage holds trivially by our definition of $q$, and Avoidance merely by using the fact that $f(1)\cap f(2)=\varnothing$.

Now we assume that $2n>m$. The construction is similar but this time it is convenient to present it in terms of the complement of each $f(i)$. Let $n'=m-n$ and $q'=\lceil \nicefrac m {n'} \rceil$. From $(k-1)m \geq nk$ we obtain $kn'\geq m$ and thus $q'\leq k$. Define
\[f(i)=Z\setminus\{((i-1)n'+j)_m\colon 1\leq j\leq n'\},\]
and as before extend $f$ to an injective function randomly. This time, Coverage holds since $f(1)\cup f(2)=Z$, whereas Avoidance holds by the way we chose $q'$.}
\vcut{
\vnote{{\color{red} VG: OK. I'd still prefer a less formal and more intuitive argument, but let us leave it there.} {\color{blue} DF: OK.}}
}
\endproof

\subsection{Short protocols for the three-agent case}
\label{SubsecShortProt}

Now we are ready to consider the general three-agent case, where Alice, Bob and Cath hold respectively $a,b,c$ cards, identified with the numbers $1,\hdots,n$ where $n=a+b+c$. Suppose the deal is $H = A \mid B \mid C$. Without loss of generality we can assume that Alice gets the first $a$ cards, Bob gets the next $b$ cards, and Cath the last $c$ cards of the deck.

To describe the protocol, first we fix a cyclic order of making announcements, e.g. first Alice, then Bob, and then Cath. When an agent gets a turn, she may make a ``real'' announcement or a ``dummy'' one, i.e. $\pass$. Suppose, without loss of generality, that Alice is the first that can  make a real announcement (we will make this precise later). Now, the protocol:

\para{Step 1} Alice chooses a card $x\not\in A$ and announces
\begin{quote}\textit{``All my cards are in the set $A' = A\cup \{x\}$}''.
\end{quote} 

\para{Step 2} Suppose the card $x$ is in Cath's hand. Then Bob passes. Note that this move tells Eaves that the extra card $x$ is not in Bob's hand, but Eaves does not know what $x$ is, so he does not know which cards of $A'$ are held by Alice and therefore safety is not violated. 

\para{Step 3} Next, Cath chooses -- if possible -- randomly a \spread
\[f_{\cat}\colon A' \rightarrow \binom {B\cup C\setminus \{x\}}{b}\]
such that $f_{\cat}(x) = B$, and makes the announcement
\begin{quote}
\textit{``For every $z\in A'$, Alice's hand is $A'\setminus \{z\}$ if and only if Bob's hand is $f_{P}(z)$''.}
\end{quote}

\para{Step 4} Finally, Alice announces \End. 
\medskip

Note that, in particular, Cath's announcement above implies that Alice's hand is  
$A'\setminus \{x\}=A$ if and only if Bob's hand is $f_{\cat}(x) = B$. Thus, the protocol is informative for each agent.  Its safety follows from the definition of \spread. We leave the details of the proof to the reader. 

\medskip
When does a \spread $f_{\cat}$ as above exist?  The conditions in Lemma \ref{lem:spread} translate as follows, for $d=b$ or $d=c$: 
\begin{enumerate}
\item $\binom {b +c -1}{d} \geq a+1$ (for injectivity). 

Assuming $b \leq c$ this becomes  $\binom {b +c -1}{c} \geq a+1$.


\item $d(a+1) \geq b+c-1$  (for coverage).  
Assuming $b \leq c$ this becomes $ba \geq c-1$.


\item $a(b+c-1) \geq d(a+1)$  (for exclusion).
Assuming $b \leq c$ this becomes $a(b-1) \geq c$, which is stronger than the inequality 2 above. 

\end{enumerate}

Note that the values of $a,b,c$ above can be permuted so as to satisfy the conditions, but once $a$ -- the number of cards of the agent who makes the first announcement -- is fixed, the conditions must hold for both cases of $d$.  

The above described protocol works for most ``balanced'' distributions, i.e. distributions where no player holds too few or too many of the cards, \vgadd{which would make the conditions of  Lemma \ref{lem:spread} unsatisfiable.} Indeed, each of the conditions in Lemma \ref{lem:spread} can be violated, so it does not cover all possible distribution types.  Here are some simple cases \vgadd{making condition 3 above unsatisfiable}: 
\begin{itemize}
\item $(1,b,c)$ for any $b,c$. 
\item $(2,b,c)$ for any $b,c$, such that $c>2b-2$. 
E.g., $(2,2,3)$, $(2,3,5)$, etc. 
\end{itemize}
\vgadd{
As examples \ref{Ex11} and \ref{Ex12} and other further results show, many of these cases are still solvable, even though not covered by Lemma \ref{lem:spread}.

\begin{example}
 \label{Ex11}

We will show the SADI problem for distribution type $(3,3,1)$ is solvable, by informally describing the following solving protocol. It is a variation of a solution to the two-agent Russian cards problem, which appeared in \cite{albertetal:2010} and, in a presentation closer to ours, in \cite{geometric}. Alice `places' all cards in the points of the 7-point projective plane, also known as the Fano plane, in such a way that her cards form a line, as indicated in Figure \ref{FigFano}. She then announces that her hand forms one of the lines. 
Without loss of generality, we can assume that the deal is $0\Comma 1\Comma 2\mid 3\Comma 4\Comma 5\mid 6$ and that she distributes the cards as shown in Figure \ref{FigFano}. In this case, she announces 

\begin{quote}
 ``My hand is one of the following: \\ $\{0\Comma 1 \Comma 2\}$, $\{0\Comma 3 \Comma 4\}$,  $\{0\Comma 5 \Comma 6\}$, $\{1\Comma 3 \Comma 5\}$, $\{1\Comma 4 \Comma 6\}$, $\{2\Comma 3 \Comma 6 \}$, or $\{2\Comma 4 \Comma 5\}$.''
\end{quote}

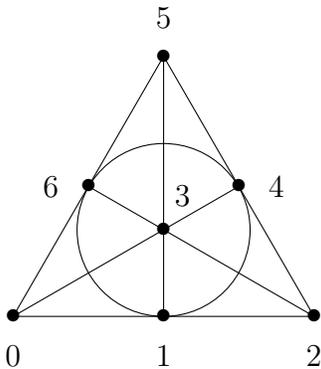
\begin{figure}[h]
\begin{center}
\begin{tikzpicture}
\node[label=below:0] (0) at (0,0) {$\bullet$};
\node[label=below:1] (1) at (2,0) {$\bullet$};
\node[label=below:2] (2) at (4,0) {$\bullet$};
\node[label=right:4] (3) at (3,1.73) {$\bullet$};
\node[label=above:5] (4) at (2,3.46) {$\bullet$};
\node[label=left:6] (5) at (1,1.73) {$\bullet$};
\node (6) at (2,1.15) {$\bullet$};
\node (6node) at (2.25,1.6) {3};
\draw (0,0) -- (4,0) -- (2,3.46) -- (0,0);
\draw (0,0) -- (3,1.73);
\draw (1,1.73) -- (4,0);
\draw (2,0) -- (2,3.46);
\draw (2,1.15) circle (1.15);
\end{tikzpicture}
\end{center}
\caption{Alice holds a line in the 7-point projective plane}
\label{FigFano}
\end{figure}

One can check by inspection on the figure that Bob immediately learns the entire deal, since there is only one line avoiding his hand. More generally, one can easily check that this would be the case whenever Alice holds a line and Bob holds three other cards.

Next, Bob must make a safe announcement, communicating to Alice and Cath each other's hands. Here we use the fact that there are seven cards (points) and seven lines, so that there is a bijection $\ell$ assigning to each point $x$ a line $\ell(x)$ such that $x$ is {\em not} on $\ell(x)$. Moreover, Bob can arrange it so that $\ell(6)=0\Comma 1\Comma 2$. He then makes an announcement consisting of a sequence of announcements of the form ``If Cath holds $x$, then Alice holds $\ell(x)$.'' One possibility is the following:
\begin{align*}
\big \{ 
\ & 
 \ 2\Comma 4\Comma 5\mid 1,3,6 \mid 0 \  ;
\ 0 \Comma 3\Comma 4\mid 2,5,6 \mid 1 \ ; 
\ 0\Comma 5\Comma 6\mid 1,3,4 \mid 2 \ ; 
\ 1\Comma 4\Comma 6 \mid 0,2,5 \mid 3 \ ; 
\\
&
\ 1\Comma 3\Comma 5\mid 0,2,6 \mid 4 \ ;  
\ 2\Comma 3\Comma 6\mid 0,1,4 \mid 5 \ :  
\ 0 \Comma 1\Comma 2 \mid 3,4,5 \mid 6 \ 
\big \}.
\end{align*}
Such announcement  is possible independently of how Alice arranges the cards on the plane. This  completes the definition of a solving protocol. 
\end{example}

\dfadd{An} interesting problem is to precisely characterize the set of solvable 3-agent \sadi problems. We will leave this for a follow-up work.}  
Our current goal, instead, is to focus on extending the techniques outlined here to cases where we have more than three agents. As it turns out, we can solve many \sadi problems with a large number of agents or cards by essentially applying the above techniques recursively in order to reduce them to simpler \sadi problems.

\section{Solvability by reduction: preliminaries and case study}
\label{SecCaseMult}
\label{SecReduct}

Here we will illustrate the method of reducing \sadi problems to simpler ones (with smaller sizes of distribution types) and eventually designing protocols for solving such problems by a sequence of such reductions.  
First, we need some preliminaries.  

\subsection{Diffusions and $k$-solvability}
\label{SecKSol}
\label{SecDiff}

The basic ideas presented in the previous section can be generalized to a larger number of agents, for which we need to make some notions precise. Cath's announcement \eqref{EqCathDiff} is a special case of a ``diffusion''. Roughly, a diffusion is a set of possible deals which, when announced, gives each of the agents enough information to fully determine the deal, but does not let the eavesdropper learn the ownership of any specific card.

\begin{definition}[Diffusion]
Fix a card distribution type $\dist$. A {\em diffusion} is a set of deals $\Delta\subseteq \deal \dist$ such that
\begin{enumerate}
\item if $H,H'\in\Delta$ are such that $H\not=H'$ and $P$ is any agent then $H_P\not=H'_P$ and
\item for every card $c\in \deck$ there are $H,H'\in \Delta$ and an agent $P$ such that $c\in H_{P}$ but $c\not\in H'_{P}$.
\end{enumerate}
 If $\#\Delta=k$, we say that $\Delta$ is a {\em $k$-diffusion} or $\Delta$ has {\em size} $k$.
\end{definition}

\vgadd{For instance, Bob's announcement in Example \ref{Ex11} is a $7$-diffusion.} In what follows, it will be very important to take into account the number of deals in a diffusion, so we introduce 
the notion of {\em $k$-solvability.} The following definition is a modification of Definition \ref{DefSolSadi}: 

\begin{definition}[$k$-solvable \sadi problems]
Let $\prsadi=(\dist,\text{\sc i},\text{\sc s})$ be a \sadi problem and let $k$ be a natural number. Say a protocol $\pi$ is a {\em $k$-solution} for $\prsadi$ if whenever $(H,\rho)$ is a terminal execution of $\pi$, then $\eavesig \rho\pi$ is a $k$-diffusion.

$\prsadi$ is {\em $k$-solvable} if it has a $k$-solution.
\end{definition}

As a ``toy case'', let us begin by studying $k$-solvability in the two-agent case. This case is, of course, trivially solvable (each agent knows the deal from the beginning so the two do not need to take any actions) but, for what will follow, we still want to know for which values of $k$ it is $k$-solvable.

\begin{lemm}\label{LemmBasic}
For any distribution type $\dist$ over two agents, Alice (\ali) and Bob (\bob), and any integer $k>1$ such that $k\leq\binom{\cardsum \dist}{s_\ali}$ and $s_\ali\leq s_\bob\leq (k-1) s_\ali$, the \sadi problem $\prsadi=(\dist, \text{\sc i},\text{\sc s})$
 is $k$-solvable.
\end{lemm}

\proof 
Let $s_\ali = a$, $s_\bob = b$, $d = a+b$  (i.e., $d= \cardsum \dist)$ and $\deck = \{1,2, \ldots d\}$. In the case of two agents, both of them know the distribution from the beginning, so no announcements are needed. Therefore all we need to show is that under the conditions of the lemma there is a $k$-diffusion $\Delta$ for the distribution type $(a,b)$. We construct it as follows. Recall that we use the notation $(x)_{d}$ to mean the unique $r\in [1,d]$ such that $x\equiv r\pmod d$ and let $m$ be the least integer such that $am \geq d$. By assumption, \vgadd{$1< m \leq k$ because $a< d$ and $ak \geq d$}. Note that a deal $H$ is uniquely determined by $H_\ali$ (since Bob holds the remaining cards), so we may define $\Delta$ in terms of Alice's hands. In the first $m$ deals in $\Delta$, Alice holds $\{(1)_{d},\ldots  (a)_{d}\}$ in the first deal, $\{(a+1)_{d},\ldots  (2a)_{d}\}$ in the second, etc., up to $\{((m-1)a+1)_{d},\ldots  (ma)_{d}\}$ in the $m$th. Thus, we ensure that every card in $\deck$ appears both in a hand of $\ali$ and in a hand of $\bob$. The remaining $(k-m)$ deals in $\Delta$, if any, we choose arbitrarily. The condition $k\leq \binom{d}{a}$ guarantees that there are at least $k$ different deals for $\dist$. 
\endproof

\subsection{A case study with multiple agents}
\label{SubsecAU}

To illustrate the notion of $k$-solvability we will outline a 
construction of a safe and informative protocol for the \sadi problem  $\prsadi$ with distribution type $(2,3,3,3)$, which will involve two recursively defined reduction steps. Let $\deck = \{0,1,\ldots, 9, 10 \}$ and let the set of agents be  $\{\ali_{0}, \ali_{1}, \ali_{2}, \ali_{3}\}$. 

\vgadd{
In the reduction techniques that we develop further for solving SADI problems, the original problem can be split into two or more sub-problems 
solved separately, and in each of these subproblems some agents may end up having no cards.} So, for technical reasons hereafter we will consider distribution types where some agents receive 0 cards, so they only occur passively. Still they are considered part of the protocol and hence they, too, hear all announcements.
Hereafter we use $\cdot$ to denote an empty hand in a deal in such type. 

\medskip
We will outline the exchange
for the deal
\[H = A_{0} |  A_{1}  | A_{2} | A_{3}=0\Comma 1| 2\Comma 3\Comma 4  | 5\Comma 6\Comma 7 | 8\Comma 9\Comma 10\]
as follows: 

\para{Step 1} The agent with 2 cards (here, agent $\ali_{0}$) chooses randomly an additional card $x_{0}$ and announces \textit{``All my cards are in the set $A^{0} = A_{0}\cup \{x_{0}\}$''.} 

\para{Step 2} Suppose without loss of generality that $x_{0}=2$, so the agent who has the card $x_{0}$ is $\ali_{1}$. Now, agent $\ali_{1}$ knows the hand of agent $\ali_{0}$ and the initial \sadi problem for the distribution type $(2,3,3,3)$, is reduced to solving the following two simpler \sadi problems:

\begin{enumerate}
\item $\prsadi_{1}$, for the distribution type $(2,1,0,0)$, including the deal  $A_{0} |x_{0}|\cdot |\cdot$. Essentially, this is a \sadi problem of type $(2,1)$ involving only the agents $\ali_{0}$ and $\ali_{1}$. It is immediately $3$-solvable, using the (only) 
$3$-diffusion
\[
\Delta_{1} = \big\{ 0,1\mid 2 \mid  \cdot \ \ ; \ \  1,2\mid 0\mid \cdot \ \ ; \ \ 2,0\mid 1\mid \cdot \big\} 
\]

\item $\prsadi_{2}$, for the distribution type $(0,2,3,3)$, including the deal 
\[\cdot| A_{1} \setminus \{x_{0}\} | A_{2} | A_{3}.\] 
\end{enumerate}

Now, the protocol essentially calls itself recursively for the \sadi problem $\prsadi_{2}$ with distribution type $(2,3,3)$, on the deal $H_{1} = A'_{1} | A_{2} | A_{3}$ where $A'_{1}  = A_{1} \setminus \{2\}$. We will trace that exchange below. 

\para{Step 2.1} Agent $\ali_{1}$ chooses randomly an additional card $x_{1}$ from the current deal $H_{1}$ and announces \textit{``All my cards, excluding the card mentioned in $A^{0}$, are in the set $A^{1} = A'_{1}\cup \{x_{1}\}$''.}

\para{Step 2.2} Suppose again w.l.o.g., that $x_{1}=5$ and hence the agent who has the card $x_{1}$ is $\ali_{2}$. Now agent $\ali_{2}$ knows the hand $A'_{1}$ of agent $\ali_{1}$ in the deal $H_{1}$ (and therefore the entire deal $H_{1}$). 
The problem $\prsadi_{2}$ is now reduced to solving the following two simpler \sadi problems: 
\begin{enumerate}
\item $\prsadi_{21}$, for the distribution type
$(2,1,0)$, including the deal  $(A'_{1} \mid \{x_{1}\} \mid \emptyset)$.  

\item $\prsadi_{22}$, for the distribution type $(0,2,3)$, including the deal $H_{2} = (\emptyset \mid  A_{2} \setminus \{x_{1}\} \mid A_{3})$. 
\end{enumerate}

This is now a base case, as both problems are immediately $3$-solvable. The only $3$-diffusion for  
$\prsadi_{21}$ is 
\[
\Delta_{21} = \big\{ 
3,4\mid 5\mid \cdot; \ 4,5\mid 3\mid \cdot; \ 5,3\mid 4\mid \cdot 
\big\}
\]
A randomly chosen 3-diffusion for $\prsadi_{22}$ involving the actual deal $H_{2}$ is e.g. $\Delta_{22} = $ 
\[\big\{ 
\cdot\mid 6,7\mid 8,9,10; \ \cdot \mid 8,9\mid 10,6,7; \ \cdot \mid 8,10\mid 6,7,9.
\big\}
\]

Now, in order for the only agent involved in both problems, $\ali_{2}$, to 
communicate the deal $H_{1}$ to $\ali_{1}$ and $\ali_{3}$, she ``fuses'' the 3-diffusions  $\Delta_{21}$ and $\Delta_{22}$ using a multi-agent analogue of a spread from Section \ref{SubsecSpread}. Namely, $\ali_2$ chooses a bijection $f\colon \Delta_{21}\to \Delta_{22}$ \vgadd{that associates the actual deals in  $\Delta_{21}$ and $\Delta_{22}$}. For example, she may define
\begin{align*}
f(3,4\mid 5\mid \cdot)&=\cdot\mid 6,7\mid 8,9,10\\
f(4,5\mid 3\mid \cdot)&=\cdot \mid 8,9\mid 10,6,7\\
f(5,3\mid 4\mid \cdot)&= \cdot \mid 8,10\mid 6,7,9.
\end{align*}
The result is the 3-diffusion

\[\Delta_{2} = 
\big\{ 3,4\mid 5,6,7\mid 8,9,10; \ 4,5\mid 3,8,9\mid 6,7,10 ; \ 3,5\mid 4,8,10\mid 6,7,9\big\}.\]
We call $\Delta_{2}$ \emph{the fusion of $\Delta_{21}$ and
$\Delta_{22}$ through $f$} and denote it  by $ \Delta_{21} \oplus_f \Delta_{22}$. 
We will define the operation $\oplus$ more formally in the next  section. 

Next, agent $\ali_{2}$ announces: 
``\textit{The deal  $H_{1}$ belongs to the set $\Delta_{2}$}''. This announcement completes the exchange for the \sadi problem $\prsadi_{2}$. It is clearly informative for all agents involved in it, i.e., $\ali_{1}$, $\ali_{2}$, $\ali_{3}$, because the first deal in $\Delta_{2}$ is the only one consistent with their hands. It is safe, too, because of the properties of diffusions. 
Indeed, every execution of the protocol for $\prsadi_{2}$ is card-safe for every card involved in $\prsadi_{2}$ because: 
\begin{itemize}
\item after the announcement of $\ali_{0}$ the eavesdropper \eaves does not learn the ownership of any card amongst
the agents $\ali_{1}$, $\ali_{2}$, $\ali_{3}$; 
\item the announcement of $\ali_{1}$ leaves each deal in $\Delta_{2}$ possible for \eaves; 
\item for every card $c$ of all those in the deal $H_{1}$ there are two deals in the diffusion $\Delta_{2}$ announced by $\ali_{2}$ which send that card in different hands. 
 \end{itemize}
 Thus, \eaves does not learn the distribution of any card in $H_{1}$. 

\para{ Step 3} Now, likewise, $\ali_{1}$, as the only agent involved in the problems $\prsadi_{1}$ and $\prsadi_{2}$, knows the entire deal $H$. In order to communicate it to the others, she constructs the fusion of the 3-diffusions $\Delta_{1}$ and $\Delta_{2}$ randomly ordered but keeping the actual deals aligned, to obtain a 3-diffusion for the original problem $\prsadi$: 
$\Delta = \Delta_{1} \oplus \Delta_{2} =$
\begin{align*}
\Big \{ \ & 0,1\mid 2,3,4\mid 5,6,7\mid 8,9,10;\\
&1,2\mid 0,4,5\mid 3,8,9\mid 6,7,10; \\
&2,0\mid 1,3,5\mid 4,8,10\mid 6,7,9 \ \Big\}.
\end{align*}
Finally, agent $\ali_{1}$ announces: 
\textit{``The deal  $H$ belongs to the set $\Delta$''.}

This completes the execution of the protocol for $\prsadi_{2}$. Again, it is clearly informative for all agents $\ali_{0}$, $\ali_{1}$, $\ali_{2}$, $\ali_{3}$, because the first deal in $\Delta$ is the only one consistent with their hands, and it is safe, because of the properties of diffusions and the construction. Indeed, we only need to note that any fusion of two $k$-diffusions of disjoint decks is a $k$-diffusion and since every deal in each of these  diffusions is possible for Eaves, so is each deal in the fusion.

As we will see in Section \ref{SecSolvability}, this is a special case of a larger class of \sadi problems which are always $k$-solvable. But first, let us give a more general theory of solvability by reduction. 

\section{Solvability by reduction: general theory}
\label{sec:reduction:general}
\label{SecReductGen}

Here we will describe the solutions presented above in a more general light. For this we need a few additional definitions and some notation.

\begin{definition}
Given distribution types $\bar s,\bar r$, we denote by $\bar s\oplus \bar r$ the standard vector sum, that is $(\bar s\oplus \bar r)_P=s_P+r_P$ for all $P\in \players$. If $S,R$ are disjoint and $H\in \deal{\bar s,S}$, $B\in \deal{\bar r,R}$, we define $A\oplus B\in \deal{\bar s\oplus \bar r,S\cup R}$ by $(A\oplus B)_P=A_P\cup B_P$. 
\end{definition}

Now we can give a generalization of {\em spread:}

\begin{definition}
Suppose that $\distr$ is a distribution type and $T,R$ are disjoint sets of cards such that $T\cup R=\deck$. Let $H$ be the actual deal and suppose that $\Gamma,\Delta$ are $k$-diffusions for $H\upharpoonright T$, $H\upharpoonright R$, respectively.
Then, a {\em spread} between $\Gamma$ and $\Delta$ is a bijection $f\colon \Gamma\to \Delta$ such that $f(H\upharpoonright T)=H\upharpoonright R$.
We also define
\[\Gamma\oplus_f\Delta=\{G\oplus f(G):G\in \Gamma\}.\]
\end{definition}

The following is very easy to check:

\begin{lemm}\label{LemmSpreadDiff}
If $f$ is a spread between $k$-diffusions $\Gamma$ and $\Delta$, then $\Gamma\oplus_f\Delta$ is a $k$-diffusion.
\end{lemm}

The following notion will be central for stating our main theorem.

\begin{definition}
Suppose that $\dist$ is a distribution type, $H$ is a deal 
of type $\dist$ and $P$ an agent. Let $T$ be a set of cards $T$ and $R=(\deck\setminus T)$ be its complement. We say $T$ is {\em splitting} (for the deal $H$ and agent $P$) if, given any deal 
$H'$ of type $\dist$ such that $H'\sim_P H$: 
\begin{enumerate}
\item there exists an agent $Q$ (possibly equal to $P$) such that $H'_Q\cap T$ and $H'_Q\cap R$ are both non-empty and
\item there exists a natural number $k$ such that $\|H'\upharpoonright T\|$ and $\|H'\upharpoonright R\|$ are both $k$-solvable. 
\end{enumerate}
\end{definition}

For example,  when Alice announces ``I hold all cards in the set $S$ but one'', then $S$ is splitting in the case of distribution type $(k-1,k,...,k)$. 
\vgadd{Indeed, the agent who holds the card in $S$ that is not in Alice's hand satisfies condition 1 above. After Alice's announcement the problem is reduced, just like in the example in Section \ref{SubsecAU}, to two sub-problems, respectively of types  $(k-1,1,0...,0)$ and $(0,k-1,k,...,k)$ of which the first is readily $k$-solvable and the second is of the same type as the original problem, but with one player less holding cards. The claim that it is $k$-solvable, too, can be proved by induction and is a particular case of the more general claim, stated and proved in Theorem \ref{TheoKNorm}.}

Observe that, in general, $S$ is splitting if and only if its complement is.
The following is trivially verified:

\begin{lemm}\label{LemmGoodPart}
If $T$ is splitting for the deal $H$ and agent $P$ and $H'\sim_P H$, then $T$ is splitting for the deal $H'$ and agent $P$.
\end{lemm}

We can now state our main reduction theorem. The strategy is to use splitting sets in order to solve \sadi problems by reducing them to simpler problems. Informally, the general idea is as follows:

\para{1} Agent $P$ chooses a splitting set $T$. Note that $T$ is also splitting for any $H'\sim_P H$ so the choice depends only on $H_P$.

\para{2} Each agent announces how many cards she holds in 
each of $T$ (and thus also in $R=\deck\setminus T$).

\para{3} \vgadd{Thus, two `sub-problems' of the original SADI problem are generated: one with $H\upharpoonright T$ and the other with $H\upharpoonright R$. The agents perform the necessary exchanges of  announcements, following the respective protocols for these sub-problems},   yielding $k$-diffusions $\Gamma$ for $H\upharpoonright T$ and $\Delta$ for $H\upharpoonright R$, respectively.

\para{4} An agent $Q$ holding a card in both $R$ and $T$ then picks a random spread $f:\Gamma\to\Delta$ and announces $\Theta=\Gamma\oplus_f\Delta$.\medskip

We formalize this in the following theorem:

\begin{theorem}\label{TheoReduct}
Suppose that $\dist$ is a distribution type such that for every deal $H$ there is an agent $P$ and a splitting set $T$ for $H$ and $P$. Then, $\prsadi=(\dist,\text{\sc i},\text{\sc s})$ is solvable.
\end{theorem}

\proof
We need to define a protocol $\pi$ which solves $\prsadi$. Suppose that the agents are numbered $P_1,\hdots, P_m$. We will define $\pi$ by describing its set of executions; since every initial segment of an execution is, by definition, an execution, it in fact suffices to describe the terminal executions. These are of the form
\[(H,\rho_0\ast\rho_1\ast\rho_T\ast\rho_R\ast\rho_2\ast \Theta\ast\End),\]
where:

\para{1}
The run $\rho_0$ has length less than $m$ where all agents pass except for $P_\ast$, who is the first agent with the property that there is a splitting set for $H$ and $P_\ast$.

\para{2}  When it is the turn of $P_\ast$, she chooses such a splitting set $T$. Define $t_P=\#(H_{P}\cap T)$ for each agent $P$. Then, in $\rho_1$, each agent $P$ (beginning with $P_\ast$) announces \textit{``I hold exactly $t_{P}$ cards from $T$''.} Note that $\rho_1$ has length exactly $m$.

\para{3} Let $R=\deck\setminus T$. By assumption there is some $k$ such that $H\upharpoonright T$ is $k$-solvable, say by a protocol $\pi_T$, as well as $H\upharpoonright R$, say by a protocol $\pi_R$. Then, $\rho_T$ is any run such that $(H\upharpoonright T,\rho_T\ast \End)$ is a terminal execution of $\pi_T$, and similarly $\rho_R$ is any run such that $(H\upharpoonright R,\rho_R\ast \End)$ is a terminal execution of $\pi_R$.

\para{4} By the definition of a splitting set there is an agent $Q_\ast$ who holds cards both in $T$ and in $R$. The run $\rho_2$ consists of less than $m$ actions where each agent who is not $Q_\ast$ passes.

\para{5} Let $\Delta\subseteq\eavesig{\pi_T}\rho$ and $\Gamma\subseteq \eavesig{\pi_R}\rho$ be $k$-diffusions and $f\colon \Gamma\to\Delta$ be a spread. The agent $Q_\ast$ announces $\Theta=\Delta \oplus _f \Gamma$. Finally, the next agent to play announces $\End$.
\medskip

We must check that  this is indeed a protocol according to our definition. For the first $m$ steps, let $\rho_0$ be an execution of less than $m$ steps of $\pi$, and suppose that it is the turn of agent $Q$. Then, if there is no splitting set for $H$ and $Q$ and $H\sim_Q H'$, then there is also no splitting set for $H'$ and $Q$ so $\pi(H,\rho_0)=\pi(H',\rho_0)=\{\pass\}$, and clearly $H\in \pass$; the situation is very similar if another agent has already made a non-trivial announcement. On the other hand, if there is a splitting set $T$ for $H$ and $P_\ast$ where $P_\ast$ is the first agent for whom this is the case, then $T$ is also a splitting set for any $H'\sim _{P_\ast} H$. Moreover, $H_{P_\ast}\cap T=H'_{P_\ast}\cap T$ so they have the same number of elements, from which it follows that $\pi(H,\rho_0)=\pi(H', \rho_0)$. Clearly $H \in \alpha$ if $\alpha$ is ``I hold $n$ cards in $T$'', where $n=\#(H_{P_\ast}\cap T)$.

Now consider an execution of the form $\rho_0\ast\rho_1$. By Lemma \ref{LemmCountAnn}, the sets $T$ and $R=\deck\setminus T$ are uniquely determined by agent $P_\ast$'s announcement, and as before if $H\sim_Q H'$ then $\#(H_Q\cap T)=\# (H'_Q\cap T)$, from which all required properties follow.

If $\rho_0\ast\rho_1\ast\rho_T$ is an execution of $\pi$ and $H\sim_Q H'$, then $H\upharpoonright T\sim_Q H'\upharpoonright T$, which means that
\[\pi(H, \rho_0\ast\rho_1\ast\rho_T)=\pi_T(H\upharpoonright T,\rho_T)=\pi_T(H'\upharpoonright T,\rho_T) = \pi(H', \rho_0\ast\rho_1\ast\rho_T),\]
and similarly from the assumption that $H\upharpoonright T\in \bigcap \pi_T(H\upharpoonright T,\rho_T)$ it follows that $H\in \bigcap \pi_T(H,\rho_T)$. Executions of the form $\rho_0\ast\rho_1\ast\rho_T\ast\rho_R$ are dealt with in a similar fashion.

If $(H,\rho_0\ast\rho_1\ast\rho_T\ast\rho_R\ast\rho_2)$ is an execution of $\pi$ and $H'\sim_P H$ is such that $(H',\rho_0\ast\rho_1\ast\rho_T\ast\rho_R\ast\rho_2)$ is also an execution of $\pi$, then $H_P$ does not intersect one of $T$ or $R$ \vgadd{(because, the agent $Q_\ast$ has not taken turn in this part of the protocol yet)} and hence $H'_P=H_P$ also does not intersect one of $T$ or $R$, hence $\pi(H,\rho_0\ast\rho_1\ast\rho_T\ast\rho_R\ast\rho')=\pi(H',\rho_0\ast\rho_1\ast\rho_T\ast\rho_R\ast\rho')=\{\pass\}$.

Finally, if $(H,\rho_0\ast\rho_1\ast\rho_T\ast\rho_R\ast\rho_2\ast\Theta)$ is an execution of $\pi$ and $H'\sim_{Q_\ast} H$ is such that $(H',\rho_0\ast\rho_1\ast\rho_T\ast\rho_R\ast\rho_2)$ is also an execution of $\pi$, then since $(H\upharpoonright T,\rho_T\ast\End)$ is a terminal run of $\pi_T$ which is informative, we have $H\upharpoonright T=H'\upharpoonright T$, and similarly $H\upharpoonright R=H'\upharpoonright R$, which means that $H=H'$ and thus $\pi(H,\rho_0\ast\rho_1\ast\rho_T\ast\rho_R\ast\rho_2)=\pi(H',\rho_0\ast\rho_1\ast\rho_T\ast\rho_R\ast\rho_2)$. Since $\Theta$ is a $k$-diffusion by Lemma \ref{LemmSpreadDiff}, it follows that this last announcement is informative to all and the following agent may announce $\End$.

It remains to check safety. Suppose that $H$ is a deal, $\rho=\rho_0\ast\rho_1\ast\rho_T\ast \rho_R\ast \Theta\ast\End$ is a run such that $(H,\rho)$ is a terminal execution of $\pi$ and $H'\in \Theta=\Gamma\oplus_f\Delta$. Then, since $\Gamma$ was a $k$-diffusion for $\pi_T$ it follows that $(H'\upharpoonright T,\rho_T)$ is an execution of $\pi_T$. Similarly, $(H'\upharpoonright R,\rho_R)$ is a run of $\pi_R$, and hence $(H',\rho)$ is also an execution of $\pi$. Since $H'\in \Theta$ was arbitrary, $\Theta\subseteq \eavesig \pi{\rho}$; safety then follows from Lemma \ref{LemmSpreadDiff} since $\Theta$ is a $k$-diffusion.
\endproof

\vgadd{
\begin{example}
 \label{Ex12}

\vcut{
\begin{figure}[h]
\begin{center}
\begin{tikzpicture}
\node[label=below:0] (0) at (0,0) {$\bullet$};
\node[label=below:1] (1) at (2,0) {$\bullet$};
\node[label=below:2] (2) at (4,0) {$\bullet$};
\node[label=right:4] (3) at (3,1.73) {$\bullet$};
\node[label=above:5] (4) at (2,3.46) {$\bullet$};
\node[label=left:6] (5) at (1,1.73) {$\bullet$};
\node (6) at (2,1.15) {$\bullet$};
\node (6node) at (2.25,1.6) {3};
\draw (0,0) -- (4,0) -- (2,3.46) -- (0,0);
\draw (0,0) -- (3,1.73);
\draw (1,1.73) -- (4,0);
\draw (2,0) -- (2,3.46);
\draw (2,1.15) circle (1.15);
\end{tikzpicture}
\end{center}
\caption{Alice holds a line in the 7-point projective plane}
\label{FigFano}
\end{figure}
}

Recall Example \ref{Ex11} where we showed that the SADI problem with distribution type $(3,3,1)$ is solvable, using the Fano plane. Since Bob's announcement there is always a $7$-diffusion, we have actually shown that it is 
$7$-solvable.

As a simple application of Theorem \ref{TheoReduct}, we can now use splittings to reduce other cases to this one. For instance, consider a distribution of type $(6,7,1)$ (again, not covered by Lemma \ref{lem:spread}). 

Here is an informal description of a solving protocol. First, Alice, who holds 6 cards, splits the deck into two subsets of 7 cards each, so that in each of them she holds three cards.  Without loss of generality, assume the deal $H$ is given by
\[0\Comma 1\Comma 2\Comma 3\Comma 4\Comma 5 \mid 6 \Comma 7\Comma 8\Comma 9\Comma 10\Comma 11\Comma 12\mid 13.\]
Then, for example, Alice splits the deck into $S=\{0\Comma 1\Comma 2\Comma 6\Comma 7\Comma 8 \Comma 13\}$ and $T= \{3\Comma 4\Comma 5\Comma 9\Comma 10\Comma 11 \Comma 12\}$. Observe that, no matter how Alice does this splitting, Cath will hold one card in one of the sets and no cards in the other, so the 
resulting distribution types of $H\upharpoonright S$ and $H\upharpoonright T$
will be $(3,3,1)$ and $(3,4,0)$ (in an unspecified order).
We already know that the SADI problem for the distribution type $(3,3,1)$ is $7$-solvable. Now, the one for distribution type $(3,4,0)$ is $7$-solvable, too. To see this, once again arrange the seven cards into the points of the Fano plane in such a way that Alice's cards form a line and let Alice announce that she holds one of the seven lines. This is already a $7$-diffusion, since Alice's hand determines the entire deal, given that there are only two agents holding cards here. 

Thus $H\upharpoonright S$ and $H\upharpoonright T$ are both $7$-solvable, so that by Theorem \ref{TheoReduct}, $(6,7,1)$ is $7$-solvable.
\end{example}
}

We will give more applications of Theorem \ref{TheoReduct} in the next section.


\section{Some general solvability theorems}\label{SecSolvability}

Here we will show that the splitting method provides solutions in a very large class of cases. This will require a more in-depth algebraic-combinatorial analysis. We will present three main results. The first two give $k$-solvability for a fixed value of $k$. Theorem \ref{TheoKNorm} may be used in many cases where the total number of cards is less than $mk^2$, although some extra assumptions are needed, including that most players have a multiple of $k$ cards. Theorem \ref{TheoBounded} shows that \sadi problems are $k$-solvable whenever no player holds too many or too few of the cards, provided the deck is large enough. Finally, Theorem \ref{TheoUnrestrict} shows that we can drop the upper bound on the number of cards a player may hold if we do not fix the value of $k$ beforehand.

We begin with two combinatorial constructions which will be useful later in this section.

\begin{lemm}\label{LemmCombinat} Let $X$ be a finite set with $n$ elements.
\begin{enumerate}
\item \label{LemmCombin1} Suppose that $a<n$ and $k>2$ are such that $n\geq k^2$ and
\[(k-1)(n-k)\leq ka\leq (k-1)n.\]
Then, there exist sets $Y_1,\hdots,Y_k$ such that for all $i\leq k$, $\# Y_i= n-a$, $\bigcap_{i\leq 3}Y_{j_i}=\varnothing$ whenever $j_1,j_2,j_3$ are all distinct, $\bigcup_{i\leq k}Y_i=X$ and $\#(Y_i\cap Y_j)\leq 2$ whenever $i\not=j$.

\item \label{LemmCombin2}
Suppose that $c n>b (b+c)$ for some natural numbers $b>c$. Then, there are sets $Y_1,\hdots,Y_k$ for some number $k$ such that $\# Y_i=b$ for all $i\leq k$, $\bigcup_{i\leq k}Y_i=X$ and $\#(Y_i\cap Y_j)\leq c+1$ whenever $i\not=j$.
\end{enumerate}
\end{lemm}

\proof 
First we prove Claim \ref{LemmCombin1}. The general idea is to arrange all $n$ elements in a rectangular table with $k$ columns and an incomplete last row. Then for each $1\leq i\leq k$ we define $Y_i$ by taking all elements in the $i$-th column plus sufficiently many from the $i$-th row to make the number of elements in $Y_i$ to be $n-a$. \vgadd{See illustration in Figure \ref{table}.}

\begin{figure}
\vgadd{
\[ 
\left.\begin{array}{|c|c|c|c|c|c|c|}
\hline
x_{1}  & \cdots & \bm{x_{i}}  & \cdots & x_{r}  & \cdots & x_{k}  \\\hline 
 x_{k+1} & \cdots & \bm{x_{k+i}}  & \cdots & x_{k+r}  & \cdots & x_{2k} \\\hline       
\cdots &  \cdots &  \cdots &  \cdots &  \cdots & \cdots & \cdots \\\hline 
 \bm{x_{(i-1)k+1}}  & \cdots & \bm{x_{(i-1)k+i}}  & \cdots & \bm{x_{(i-1)k+r}}  & \cdots & \bm{x_{ik}}  \\\hline    
  \cdots &  \cdots &  \cdots &  \cdots &  \cdots & \cdots & \cdots \\\hline  
       x_{qk+1}  & \cdots & \bm{x_{qk+i}}  & \cdots & x_{qk+r} &   &  \\\hline   
\end{array}\right.
\]
\caption{Selection of the set $Y_{i}$ (the case when $i<r$).}
\label{table}
}
\end{figure}

For the technical details, write $n=qk+r$ with $0\leq r < k$. Note that $q\geq k$.  
Consider the set
\[I=\{(i,j): 1\leq i\leq k\text{ and }1\leq j \leq q\text{ or }1\leq i\leq r\text{ and }j=q+1\};\]
it has $n$ elements, so we may use $I$ to enumerate $X$, and write $X=\{x_{ij}:(i,j)\in I\}$.

Next we claim that $a\leq n-q$ if $r=0$ and $a\leq n-(q+1)$ if $r\not =0$. In the first case, we have that $ka\leq (k-1)n=kn-n=kn-kq$, so $a\leq n-q$. In the second, $ka\leq (k-1)n=kn-kq-r$ and $0<r<k$ so $a\leq n-q-1$. 

Furthermore, from $(k-1)(n-k)\leq ka$ we get $kn-ka\leq n+k^2-k$, hence $n-a\leq \frac{n}{k}  +k-1$, so  $n-a\leq \lfloor \frac{n}{k} \rfloor  +k-1$ = $q+k-1$. 

Now, for $1\leq i\leq k$ we define $Y_i$ as follows.
First, let $Y'_i$ be the set of all elements of $X$ of the form $x_{ij}$;
observe that each $Y'_i$ will have either $q$ or $q+1$ elements. Thus,
\begin{equation}\label{EqBoundY}
(n-a)-\#Y'_i\leq (n-a)-q\leq k-1.
\end{equation}
Given a fixed $i$, there are then exactly $k-1$ elements of the form $x_{ji}$ 
with $j\not=i$. 
We will choose $Y''_i$ from among these elements in such a way that $Y_i=Y'_i\cup Y''_i$ has exactly $n-a$ elements, 
which is possible in virtue of \eqref{EqBoundY}. It is then straightforward to check that, if $i\not=j$, $Y_i\cap Y_j\subseteq \{x_{ij},x_{ji}\}$, and thus the sets $Y_1,\hdots,Y_k$ satisfy all desired properties.

The proof of Claim \ref{LemmCombin2} uses a similar idea. Write $n=qb+r$ with $r<b$ and then write the elements of $X$ as $x_{ij}$ where either $1\leq i\leq b$ and $1\leq j\leq q$ or $1\leq i\leq r$ and $j=q+1$. Then, for $i\leq q$ let $Y_i$ be the set of all $x_{ij}$. If $r=0$, we are done, otherwise let $Y_{q+1}$ be chosen as follows. Let $Y'_{q+1}$ be those elements of the form $x_{i,q+1}$. Then choose $Y''$ to be an arbitrary $(b-r)$-element subset of the set
\[G=\{x_{ij}:i\in[1,c]\text{ and }j\leq q\};\]
it is possible to select such a $Y''$ since
\[c qb+c b>c qb+c r=cn>b(b+c) =b^2+c b, \]
so that $c qb>b^2$ and thus $\#G=c q>b$.

Then set $Y_{q+1}=Y'_{q+1}\cup Y''_{q+1}$. It is easy to check that the sets $Y_1,\hdots,Y_{q+1}$ have the desired properties.
\endproof

\subsection{Solvability in relatively small cases}

We may solve many \sadi problems with a relatively small number of cards, but we will need a few conditions on how the cards are distributed. Distribution types satisfying such conditions will be called {\em $k$-normal.}

\begin{definition}
\label{def:k-normal}
A distribution type is $k$-normal if there are at least two agents, and there is an agent $\ali$ such that
\begin{enumerate}
\item $s_\ali\equiv -1\pmod k$
\item if $P\not=\ali$, $s_P\equiv 0\pmod k$
\item if $P$ is any agent, $s_P\leq (k-1)^2$.
\end{enumerate}
\end{definition}

\begin{theorem}\label{TheoKNorm}
Given \vgadd{$k > 2$} and any $k$-normal distribution $\dist$, the \sadi problem $(\dist,\text{\sc i,s})$ is $k$-solvable.
\end{theorem}

\proof
\vgadd{Note that the number of cards in any $k$-normal distribution $\dist$ equals $qk-1$ for some $q\geq 2$. We proceed by induction on $q$. The base case, when $q=2$ follows immediately from Lemma \ref{LemmBasic}. Suppose that the claim holds for some $q\geq 2$, for any number (at least 2) of agents and every $k$-normal  distribution with $qk-1$ cards. Now, let  $\dist$ has $(q+1)k-1$ cards.} 
Consider two cases. 

\para{1} If there are only two agents, say with $s_1\leq s_2$, then we have that $s_1\geq k-1$ whereas $s_2\leq (k-1)^2$. It then follows from Lemma \ref{LemmBasic} that the \sadi problem is $k$-solvable. \vgadd{Note that this argument does not depend on which of the two agents has `$-1$ modulo $k$'  many cards.} 

\para{2} Otherwise, Alice (\vgadd{the agent $\ali$ from Definition \ref{def:k-normal}}) chooses at random a set of $k-1$ cards that she holds (say, $A$) and one that she does not (say, $b$) and announces that she holds $k-1$ cards from $A\cup \{b\}$. Once again by Lemma \ref{LemmBasic}, the \sadi problem $(\dist\upharpoonright A\cup\{b\},\text{\sc i,s})$ is $k$-solvable. Meanwhile, observe that in $\dist\upharpoonright(\deck\setminus (A\cup \{b\}))$, Alice now holds a multiple of $k$ cards, whereas the unique agent who holds $b$ now has $-1$ cards modulo $k$, 
\vgadd{so this is a $k$-normal distribution with $qk-1$ cards.} 
It follows from the induction hypothesis that $(\dist\upharpoonright(\deck\setminus (A\cup \{b\}),\text{\sc i,s})$ is $k$-solvable and, by Theorem \ref{TheoReduct}, so is the \sadi problem $(\dist,\text{\sc i,s})$, as claimed.
\endproof

As an example, consider the distribution type $\dist=(5,12,18,24)$. \vgadd{(For convenience, we have listed the sizes of the hands in an increasing order.)} All agents hold a multiple of $6$ cards, except for the first whose number of cards is $-1$ modulo $6$. Besides, $(6-1)^2=25$, and no agent holds more than $25$ cards. It follows \vgadd{by Theorem \ref{TheoKNorm}} that $\dist$ is $6$-solvable. More generally, we may consider a distribution of the form $(x_0(k-1),x_1k,x_2k\hdots,x_m k)$ with $m+1$ agents and each $x_i\leq k-2$, so that $x_ik< (k-1)^2$.  By Theorem \ref{TheoKNorm}, the \sadi problem for such a distribution is always $k$-solvable.

\subsection{Bounded solvability theorem}
\label{SubsecBalSolvability}

With larger decks, we may dispense with the assumption that most players hold a multiple of $k$ cards. Here we will present a general solvability result which essentially claims that the \sadi problem $(\dist,\text{\sc i},\text{\sc s})$ is solvable for all large enough and `sufficiently balanced' distributions with both lower and upper bounds on the size of each individual hand. The general strategy will be to `unbalance' the distribution by taking cards away from all players but Alice, until she holds a fairly large portion of the cards so that we may apply the following result.

\begin{lemm}\label{LemmAlmostLops}
If $\dist$ is a distribution type such that $|\dist|\geq k^2$, each player has at least three cards and
\[(k-1)(|\dist|-k)\leq k s_\ali\leq (k-1)|\dist|,\]
then $\dist$ is $k$-solvable.
\end{lemm}

\proof
\vgadd{
The protocol goes as follows. 
First, Alice chooses sets $Y_1,\hdots Y_k$ as in Lemma \ref{LemmCombinat}, such that $\deck\setminus H_\ali=Y_i$ for some $i$ (the latter condition may be enforced by choosing an appropriate permutation) and announce that her hand is one of the $\deck\setminus Y_i$.
All players then know Alice's cards because every player holds at least three cards and therefore can distinguish between the $Y_i$'s, since every two of these have at most 2 cards in common.

Then, every other player announces in turn} ``If Alice holds $\deck \setminus Y_i$ then my hand is $A_i$'', where $A_i$ is their true hand when $Y_i=\deck\setminus H_\ali$ and always $A_i\subseteq Y_i$. These announcements can be chosen randomly, provided they  \vgadd{do not contradict the previous players' announcements}. More precisely, we introduce auxiliary sets $Z_1,\hdots Z_k$ which we initialise after Alice's announcement as $Z_{i} = Y_{i}$, for each $i = 1,\hdots k$. Then every next player chooses for each $i = 1,\hdots k$ a subset $A_i$ of $Z_i$ of size equal to the number of cards in that player's hand and makes the announcement 
\begin{quote}
If Alice holds $\deck \setminus Y_1$ then my hand is $A_1$, if Alice holds $\deck \setminus Y_2$ then my hand is $A_2$, \textellipsis and if Alice holds $\deck \setminus Y_n$ then my hand is $A_n$.
\end{quote}
In case when $i$ is such that $Y_i=\deck\setminus H_\ali$, the player makes the only possible truthful announcement by choosing $A_i$ to be her hand. After every such  announcement, the set $Z_i$ is updated by removing the elements of $A_i$ from it. 

An easy inductive argument shows that every player has a choice of correct announcement for each $i$ and that the safety of the protocol is preserved. The latter follows from the choice of initial announcement of Alice.
\endproof

Of course, we are interested in a much more general class of distribution types, but Lemma \ref{LemmAlmostLops} will be very useful since we may reduce many other distributions to ones where a player holds most of the cards. The following will be the more technical presentation of this idea, but later we will give easier bounds to show its scope.

\begin{lemm}\label{LemmLessBig} 
Consider a \sadi problem $(\dist,\text{\sc i},\text{\sc s})$
with $m$ players and suppose that Alice holds at least $\frac{|\dist|}m$ cards and $k\geq 4$ is such that for $d=\frac{(k-1)|\dist|-ks_\ali}{k^2-3k+1}$ we have
\begin{enumerate}
\item for each agent $P$, $ks_P\leq (k-1)|\dist|$,
\item $(2k-1)(m-1)<|\dist|-s_\ali-d(k-2)$,\label{ItemPigeon}
\item $|\dist|-dk\geq k^2$ and
\item each player holds at least $k$ cards except possibly for one who holds exactly $k-1$ cards.
\end{enumerate}
Then $\dist$ is $k$-solvable.
\end{lemm}

\proof[Proof sketch]
We use complete induction on $|\dist|$: assuming the claim holds for all distributions with lesser size we will show that it holds for the given size.

Consider two cases. If for some $P$ we have that $(k-1)(|\dist|-k)\leq ks_P$, 
then we may use Lemma \ref{LemmAlmostLops} directly. Otherwise, 
we choose a player $P$ as follows. If one player has $k-1$ cards, this player is $P$. If not, due to item \ref{ItemPigeon} and the pigeonhole principle, there must always be a player $P$ different from Alice with at least $2k-1$ cards. That player announces $k$ cards out of which she holds $k-1$. Let $\dist'$ be the remaining distribution; note that $|\dist'|=|\dist|-k$. We must check, using the induction hypothesis,  that each condition still holds for $\dist'$ and $d'=\frac{(k-1)|\dist'|-ks'_\ali}{k^2-3k+1}$. This boils down to fairly standard algebraic manipulations which we have included in the appendix.
\endproof

As a direct application we obtain the following solvability result; the proof will also be left for the appendix.

\begin{theorem}[Restricted solvability] \label{TheoBounded}
Given $m>2$ and $k>2m$ there exists $N$ such that whenever $\dist$ is a distribution for $m$ players such that $|\dist|>N$ and for each $P$, $k^2\leq k s_P \leq (k-1)|\dist|,$ then $\dist$ is $k$-solvable.
\end{theorem}

\subsection{Unrestricted solvability theorem}
\label{SubsecUnrestSolvability}

We will now turn to proving a version of the previous result which implies solvability of all large enough and `semi-balanced' distributions, without  prescribing a value of $k$ and without imposing upper bounds on the size of the individual hands. We first state the result, but the proof will require several steps.

\begin{theorem}[Unrestricted solvability]\label{TheoUnrestrict}
Given $m$ there is $N$ such that whenever $|\dist|>N$ is a distribution over at most $m$ players and each player holds at least $\frac 12\sqrt{\nicefrac{|\dist|}m}$ cards then $(\dist,\text{\sc s},\text{\sc i})$ is solvable.
\end{theorem}

We will split the proof into two cases, each covered by a separate lemma. The first is analogous to Theorem \ref{TheoBounded}, except that the value of $k$ now depends on $|\dist|$. We defer the proof to the appendix.

\begin{lemm}\label{LemmSmallA}
Given $m$ there exists $N$ such that whenever $\dist$ is a distribution for $m$ players such that $|\dist |>N$ and for each $P$,
\[\frac 12\sqrt{\nicefrac{|\dist|}m} \leq s_P \leq |\dist|-2m\sqrt{|\dist|},\]
then $(\dist,\text{\sc s},\text{\sc i})$ is solvable.
\end{lemm}

We consider the case where one player holds a very large portion of the deck separately.

\begin{lemm}\label{LemmBigA}
If $\dist$ is any distribution type such that each player has more than $8m^2$ cards, $s_\ali\geq|\dist|-2m\sqrt{|\dist|}$ and $|\dist|$ is large enough then $\dist$ is solvable.
\end{lemm}

\proof
Set $n=|\dist|$ and $b=|\dist|-s_\ali\leq 2m\sqrt n $. We will use Lemma \ref{LemmCombinat}.\ref{LemmCombin2}, from which the condition $c n>b( b+c)$ is equivalent to $c>\frac{b^2}{n-b}$, so it is sufficient to have
\[c>\frac{4m^2n}{n-2m\sqrt n}.\]
For large $n$, it suffices to set $c=\frac{4m^2n}{\nicefrac n2}=8m^2$.

Then, Alice may choose sets $Y_1,\hdots,Y_k$ satisfying the conditions of Lemma \ref{LemmCombinat}.\ref{LemmCombin2} and such that $\deck\setminus H_\ali=Y_{i_\ast}$ for some $i_\ast$ (the latter condition is obtained by permuting the cards appropriately). She then announces that she holds one of the $\deck\setminus Y_i$, after which each other player $P$ knows the deal since $Y_i\cap Y_j$ can have only $8m^2$ elements and thus $H_P$ is contained in a single $Y_i$. The other players then make an announcement of the form {\em If my cards are contained in $Y_i$, then I hold $A_i$,} where $A_i\subseteq Y_i$ is chosen at random except that $A_{i_\ast}=H_P$. Observe that the players must make announcements which are consistent with each other, but as we have seen before, this is easy to accomplish provided they make their announcements one at a time.
\endproof

With these results, Theorem \ref{TheoUnrestrict} becomes immediate.

\proof[Proof of Theorem \ref{TheoUnrestrict}]
We consider two cases. If $s_\ali\leq |\dist|-2m \sqrt {\cardsum{\dist}}$, we apply Lemma \ref{LemmSmallA}. If  $s_\ali\geq |\dist|-2m\sqrt{|\dist|}$ we apply Lemma \ref{LemmBigA}, taking $|\dist|$ large enough so that $\frac 12\sqrt{\nicefrac{|\dist|}m}>8m^2$.
\endproof

\section{Concluding remarks}
\label{SecConcl}

We have introduced and studied a generic problem about secure exchange and aggregation of distributed information in multi-agent systems by public announcements, presumed intercepted by an eavesdropper.  We are interested in \dfadd{{\em unconditional}} information security, based not on encrypting that is computationally hard to break but on the combinatorial properties of the protocols.  

We have modelled and formalised the general Secure Aggregation of Distributed Information (SADI) problem as a multi-agent generalization and modification of the Russian cards problem. As we have seen, such a generalization gives rise to some issues that were not present in the original problem. One of them is that there is more flexibility in the notions of security and informativity that may be considered. Here we have focused on card-safe, informative protocols, but other combinations may also be of interest.

We note that, since we consider more than two agents, the problem is still quite non-trivial even though the eavesdropper holds no cards. Still, we have developed some general techniques for designing safe and informative protocols and have obtained \dfadd{computable} solutions for a large class of SADI problems, covering all large enough and sufficiently balanced distributions. 

It should be noted that, while Theorem \ref{TheoReduct} works for any splitting set $T$, \dfadd{in our main applications} we only used the special case where $T$ was of the form $A\cup \{x\}$, where, if $P$ is the agent making the announcement, then $A\subseteq H_P$; in other words, agents only choose one card they do not hold when splitting. However, in future work we plan on extending the applications using a wider class of splitting sets, \dfadd{such as those of Example \ref{Ex12}, in order to solve a wider class of \sadi problems.} Eventually, we hope to obtain a complete classification of all \sadi problems of the type considered here into solvable or unsolvable, and to develop sufficiently strong techniques to design solutions to all solvable cases.

\dfadd{Finally, we note that we have only considered \vgadd{card-safe} \sadi problems, for which as we have shown to have the benefit that they are solvable provided they satisfy some very mild conditions. However, for many practical applications, the more stringent notion of  \vgadd{\em strongly card-safe} security might be desirable, which will be studied in future work. Eventually, we hope and expect that our results and methods can be applied to developing practically useful secure communication protocols; in particular, for design and secure exchange of sensitive information, such as passwords, bank details, private RSA keys, etc. between distributed agents over insecure channels.}

\begin{appendix}
\section{Technical proofs}\label{Appdx}

In this Appendix we include the proofs of Theorems \ref{TheoBounded}, Lemma \ref{LemmLessBig} and Lemma \ref{LemmSmallA}.

\proof[Proof details for Lemma \ref{LemmLessBig}] We continue using the assumptions and notation from the proof sketch; recall that it only remained to check the case where for all players $Q$, $ks_Q< (k-1)(|\dist|-k)$. Recall also that we had chosen an agent $P$ such that either she holds $k-1$ cards if such a player exists, or she holds at least $2k-1$ cards. That player announces $k$ cards out of which she holds $k-1$. Let $\dist'$ be the remaining distribution, so that $|\dist'|=|\dist|-k$. We must check that Conditions 1--4 still hold for the new distribution.

For Condition 1 we have that, since we had $ks_Q <(k-1)(|\dist|-k)$ for all $Q$, we now have $k\dist'_Q<(k-1)(|\dist|-k)=(k-1)|\dist'|$. Condition 4 holds since either $P$ holds at least $k$ cards and all other players hold at least $k$ cards as well except for possibly a single other player who holds $k-1$. The exception to this is when $P$ held $k-1$ cards, but in this case she holds no cards in the new subproblem and hence does not participate in the exchange.

For Conditions 2 and 3 we must consider two subcases. It may be that from the $k$ cards announced by player $P$, Alice holds the card that $P$ did not hold, in which case $s'_\ali=s_\ali-1$, or that a different player holds it and $s'_\ali=s_\ali$. In both cases we must check that each condition still holds for $\dist'$ and $d'=\frac{(k-1)|\dist'|-ks'_\ali}{k^2-3k+1}$ in order to apply the induction hypothesis.

First assume that Alice holds the remaining card. Condition 2 holds since
\begin{align*}
(2k-1)&(m-1)<|\dist|-s_\ali-d(k-2) \\
=& |\dist|-s_\ali-\left(\frac{(k-1)|\dist|-ks_\ali}{k^2-3k+1}\right)(k-2)\\
= &(|\dist'|+k)-(s'_\ali+1)-\left(\frac{(k-1)(|\dist'|+k)-k(s'_\ali+1)}{k^2-3k+1}\right)(k-2)\\
= &|\dist'|-s'_\ali-d'(k-2)-\frac 1{k^2-3k+1}\\
<&|\dist'|-s'_\ali-d'(k-2),
\end{align*}
and Condition 3 because
\begin{align*}
k^2 &\leq |\dist|-dk
= |\dist|-\left(\frac{(k-1)|\dist|-ks_\ali}{k^2-3k+1}\right)k\\
&= |\dist'|+k-\left(\frac{(k-1)(|\dist'|+k)-k(s'_\ali+1)}{k^2-3k+1}\right)k\\
&= |\dist'|-d'k-\left(\frac{k-1}{k^2-3k+1}\right)k< |\dist'|-d'k.
\end{align*}
Thus Conditions 1--4 all hold and we may use our induction hypothesis to see that $\dist'$ is $k$-solvable.

Now we consider the case where $s'_\ali=s_\ali$, and proceed with checking Conditions 2 and 3 once again. The argument is very similar; in this case for Condition 2 we have that
\[(2k-1)(m-1)<|\dist|-s_\ali-d(k-2)= |\dist'|-s'_\ali-d'(k-2)-\frac {k}{k^2-3k+1}
\]
so $(2k-1)(m-1)<|\dist'|-s'_\ali-d'(k-2)$, whereas we obtain Condition 3 from
\begin{align*}
k^2 \leq |\dist|-dk &= |\dist'|+k-\left(\frac{(k-1)(|\dist'|+k)-ks'_\ali}{k^2-3k+1}\right)k \\
&= |\dist'|-d'k-\left(\frac{2k-1}{k^2-3k+1}\right)k,
\end{align*}
and thus $k^2 <|\dist'|-d'k$. So in either case, $\dist'$ is $k$-solvable by the induction hypothesis; since its complement is also $k$-solvable by Lemma \ref{LemmBasic} (because Alice holds $k-1$ cards and a single other player holds one), it follows that $\dist$ is solvable by Theorem \ref{TheoReduct}.
\endproof

Theorem \ref{TheoBounded} and Lemma \ref{LemmSmallA} are corollaries of this general result, but before we proceed, let us establish two bounds which will be useful below.

\begin{lemm}\label{LemmSimpBound}
Suppose that $a,k,m,n$ are positive integers such that $\nicefrac nm\leq a$ and let $d=\frac{(k-1)n-ak }{k^2-3k+1}$. Then,
\begin{enumerate}
\item\label{LemmSimpBoundOne} $\frac{n(k-m-1)}{m(k^2-3k+1)} \leq  n-a- d(k-2)$ and
\item\label{LemmSimpBoundTwo} $n\left(\frac{k(\frac{k}m-2)+1}{k^2-3k+1}\right) \leq n-dk$.
\end{enumerate}
\end{lemm}

\proof
For the first claim, we see that $\frac{n(k-m-1)}{m(k^2-3k+1)} \leq   \frac{a (k-1)-n}{k^2-3k+1}$ by writing $\frac{n(k-m-1)}{m(k^2-3k+1)}=\frac{\frac nm (k-1)-n}{k^2-3k+1}$ and using the assumption that $\nicefrac nm\leq a$. But plugging in values and simplifying, we obtain $\frac{a (k-1)-n}{k^2-3k+1} =n-a- d(k-2)$.

For the second, plugging in values and simplifying we see that $n-dk= \frac{k^2a-(2k-1)n}{k^2-3k+1}$; but once again we use the fact that $\nicefrac nm\leq a$ to obtain
\[n\left(\frac{k(\frac{k}m-2)+1}{k^2-3k+1}\right) = \frac{k^2\left(\frac nm\right)-(2k-1)n}{k^2-3k+1} \leq \frac{k^2a-(2k-1)n}{k^2-3k+1}.\qedhere\]
\endproof

Now we are ready for the final two proofs.

\proof[Proof of Theorem \ref{TheoBounded}]
We assume, as in the statement of the theorem, that $m>2$, $k>2m$, $\dist$ is a distribution for $m$ players and for each $P$, $k^2\leq k s_P \leq (k-1)|\dist|$. The result will follow from Lemma \ref{LemmLessBig} if we show that Conditions 1--4 hold when $|\dist|$ is large. By the pigeonhole principle, the player with most cards (which we may assume to be Alice) has at least $\lceil\nicefrac{|\dist|}m\rceil$ cards. Let $n=|\dist|$ and $a=s_\ali$ and recall that $d=\frac{(k-1)n-ak }{k^2-3k+1}$.

\para{Condition 1} We have by assumption that $k s_P \leq (k-1)|\dist|$ for every player $P$.

\para{Condition 2} Since $k>2m > m+1$ we have that $k-m-1>0$. Moreover, since $m>2$ it follows that $k>4$, and one can easily check that $k^2-3k+1>0$. Thus $\frac{(k-m-1)}{m(k^2-3k+1)}$ is positive, which implies that $(2k-1)(m-1)\leq {\frac{n(k-m-1)}{m(k^2-3k+1)}}$ for large $n$, since the left-hand side is fixed, whereas the right-hand side is linearly increasing on $n$. We may then use Lemma \ref{LemmSimpBound}.\ref{LemmSimpBoundOne} to obtain $(2k-1)(m-1)\leq  n-a- d(k-2)$ for large enough $n$.

\para{Condition 3} By Lemma \ref{LemmSimpBound}.\ref{LemmSimpBoundTwo}, $n\left(\frac{k(\frac{k}m-2)+1}{k^2-3k+1}\right)\leq n-dk.$ Since by assumption $k>2m$, $\frac{k}m-2>0$; since, also by assumption, $m>2$, we have $k>4$ which implies that $k^2-3k+1>0$. Thus for large $n$ we obtain $k^2< n\left(\frac{k(\frac{k}m-2)+1}{k^2-3k+1}\right)$, since once again the left-hand side is fixed but the right-hand side is increasing on $n$. It follows that $k^2<n-dk$, as needed.

\para{Condition 4} Each player holds at least $k$ cards by assumption.
\medskip

Thus for large enough $n$ we may apply Lemma \ref{LemmLessBig} and conclude that $(\dist,\text{\sc s},\text{\sc i})$ is solvable.
\endproof

\proof[Proof of Lemma \ref{LemmSmallA}]
The proof is very similar to that of Theorem \ref{TheoBounded}. Assume that for each $P$, $\frac 12\sqrt{\nicefrac{|\dist|}m} \leq s_P \leq |\dist|-2m\sqrt{|\dist|};$ once again, the player with most cards (which we may assume to be Alice) has at least $\lceil\nicefrac{|\dist|}m\rceil$ cards.

Set $n=|\dist|$, $a=s_\ali$, $k=\left\lceil\frac{\sqrt n}{2m}\right \rceil$ and $d=\frac{(k-1)n-ak}{k^2-3k+1}$. We will show that all conditions of Lemma \ref{LemmLessBig} hold.

\para{Condition 1} Multiplying the inequality 
$s_P \leq n-2m\sqrt{n}$
on both sides by $k=\left\lceil\frac{\sqrt n}{2m}\right \rceil$ we obtain
\[ks_P\leq k(n-2m\sqrt n)
\leq \left (k-\frac{\left( \frac{\sqrt n}{2m} \right )2m\sqrt n}n\right)n=(k-1)n\]
for every player $P$ by assumption and our definition of $k$.

\para{Condition 2} We have that $(2k-1)(m-1)\leq 2\left(\frac{\sqrt n}{2m}+1\right)m= \sqrt {n}+2m
,$ whereas by Lemma \ref{LemmSimpBound}.\ref{LemmSimpBoundOne},
\[n-a-d(k-2)\geq \frac{n(k-m-1)}{m(k^2-3k+1)}=\frac{n\left (\frac{\sqrt n}{2m}\right)}{m\left (\frac{n}{4m^2}\right)}+o(\sqrt n)=2\sqrt n+o(\sqrt n).\]
Thus for large $n$, $(2k-1)(m-1)< n-a-d(k-2).$

\para{Condition 3} Observe that from $k=\left\lceil\frac{\sqrt n}{2m}\right \rceil\leq \frac{\sqrt n}{2m}+1$ we obtain $k-1 \leq \frac{\sqrt n}{2m}$ and hence $4m^2(k-1)^2 < n$. Thus by Lemma \ref{LemmSimpBound}.\ref{LemmSimpBoundTwo},
\[n-dk > 4m^2(k-1)^2\left(\frac{k(\frac{k}m-2)+1}{k^2-3k+1}\right)=4mk^2+o(k^2).\]
It follows that for large $n$, $k^2<n-dk$.

\para{Condition 4} Each player holds at least $k$ cards by assumption.
\medskip

Having established Conditions 1--4, once again the result is immediate by Lemma \ref{LemmLessBig}.
\endproof
\end{appendix}

\section*{Acknowledgements}
\dfadd{To be added later.}

\vcut{
\vgadd{We are very grateful to the anonymous reviewer for the meticulous and benevolent reading of the original submission and the numerous corrections and suggestions that have significantly contributed to the improvement of the paper. The research of David Fern\'{a}ndez-Duque was supported by the {\em Ministerio de Econom\'ia y Competitividad de Espa\~na} grant number FFI2011-15945-E. 
Part of the work of Valentin Goranko was done during his visit in 2013 as a scientific expert to the Centre International de Mathematiques et d'Informatique (CIMI) Excellence Program of the University of Toulouse.}
}

\bibliographystyle{plain}

\end{document}